# AIS-MACA- Z: MACA based Clonal Classifier for Splicing Site, Protein Coding and Promoter Region Identification in Eukaryotes

*Pokkuluri Kiran Sree[1]\*, Inampudi Ramesh Babu[2] , SSSN Usha Devi N [3]*
[1]Department of CSE, JNTU Hyderabad, India
[2]Department of CSE, ANU, Guntur, India
[3]Department of CSE, University College of Engineering, JNTUK, India

*Abstract*

*Bioinformatics incorporates information regarding biological data storage, accessing mechanisms and presentation of characteristics within this data. Most of the problems in bioinformatics and be addressed efficiently by computer techniques. This paper aims at building a classifier based on Multiple Attractor Cellular Automata (MACA) which uses fuzzy logic with version Z to predict splicing site, protein coding and promoter region identification in eukaryotes. It is strengthened with an artificial immune system technique (AIS), Clonal algorithm for choosing rules of best fitness. The proposed classifier can handle DNA sequences of lengths 54,108,162,252,354. This classifier gives the exact boundaries of both protein and promoter regions with an average accuracy of 90.6%. This classifier can predict the splicing site with 97% accuracy. This classifier was tested with 1, 97,000 data components which were taken from Fickett & Toung , EPDnew, and other sequences from a renowned medical university.*

*Keywords*: *MACA(Multiple Attractor Cellular Automata), CA(Cellular Automata), AIS (Artificial Immune System), Clonal Algorithm, AIS-MACA-Z(Artificial Immune System- Multiple Attractor Cellular Automata-Version Z)*

\**Author for Correspondence* E-mail profkiransree@gmail.com

## INTRODUCTION

In recent years, study of Cellular Automata (CA) as a potential modeling tool has gained importance. Some researchers and scientists have used CA in image processing, data compression, pattern recognition, encryption, VLSI design and language recognition. Cellular Automata (CA) is a computing model which provides a good platform for performing complex computations with the available local information. CA is portrayed by local interconnectivity of cells in the network/grid. The interactions/communications between the cells are pulley local. Each cell is permitted to interact with its neighboring cells only. Further, the interconnection links typically convey just a little measure of data. No cell in the entire network will have the global view. These characteristics of CA attracted us to propose a classifier which can be very much useful for solving many problems in bioinformatics with the existing frame work.

Artificial Immune System is a novel computational intelligence technique with features like distributed computing, fault /error tolerance, dynamic learning, adaption to the frame work, self monitoring, non uniformity and several features of natural immune systems. AIS take its motivation from the standard immune system of the body to propose novel computing tools for addressing many problems in wide domain areas. These features of AIS are used in the thesis to strengthen the proposed CA classifier

## Literature Survey
Vitoantonio Bevilacqu [1] et al. tried to provide theoretical foundations for solving some problems in bioinformatics using artificial immune system like multiple sequence alignment problem and protein structure prediction. Hybrid immune algorithm was proposed for addressing multiple sequence alignment problems. Some open problems in bioinformatics are discussed and authors tried to create insight for applying AIS in bioinformatics. Shane Dixon at al has proposed Bioinformatics data mining was proposed with AIS and Neural Network. Variations in the real valued negative selection algorithm and multi layer feed forward neural network model are discussed in detail.

Niloy Ganguly [2] at al has made a survey on cellular automata which say CA uses the local information and performs complex computations. Authors gave a brief discussion on the types of Cellular Automata. Niloy Ganguly at al has also proposed theoretical concept of proposing CA for pattern classification which can be applied for low cost VLSI implementation. This classifier is capable of accommodating noise based on distance metric also. Palsh Sarkar [3]also have given a brief history of cellular automata regarding the way for creating CA games like game of life and firing squad problem and creating local CA rules for specific problems. Pradipta Maji [4] at al has proposed the error correcting capability of cellular automata based on associative memory. The desired CA is evolved with formulation of simulated annealing program. Xiao[5] at al has used CA to generate image representation for biological sequences. The research is amide to improve the quality of predicting protein attributes such as structural class and sub cellular location. Adriana Popovici at al has successful applied CA in image processing. Parallelism in CA is used to remove the noise and detection of boarders in digital images.

Jesus P. Mena-Chalco [6] at al has used Modified Gabor-Wavelet Transform for addressing this issue. In this connection, numerous coding DNA model-free systems dependent upon the event of particular examples of nucleotides at coding areas have been proposed. Regardless, these techniques have not been totally suitable because of their reliance on an observationally predefined window length needed for a nearby dissection of a DNA locale. Authors present a strategy dependent upon a changed Gabor-wavelet transform for the ID of protein coding areas. This novel convert is tuned to examine intermittent sign parts and presents the focal point of being free of the window length. We contrasted the execution of the MGWT and different strategies by utilizing eukaryote information sets. The effects indicate that MGWT beats all evaluated model-autonomous strategies regarding ID exactness. These effects demonstrate that the wellspring of in any event some piece of the ID lapses handled by the past systems is the altered working scale. The new system stays away from this wellspring of blunders as well as makes an instrument accessible for point by point investigation of the nucleotide event

Changchuan Yin [5] at el has proposed a strategy to foresee protein coding areas is produced which is dependent upon the way that the vast majority of exon arrangements have a 3-base periodicity, while intron groupings don't have this interesting characteristic. The technique registers the 3-base periodicity and the foundation clamor of the stepwise DNA sections of the target DNA groupings utilizing nucleotide circulations as a part of the three codon positions of the DNA successions. Exon and intron successions might be recognized from patterns of the degree of the 3-base periodicity to the foundation commotion in the DNA groupings.

### Design of AIS-MACA-Z
The general design of AIS-MACA-Z is indicated in the figure 1. Input to AIS-MACA-Z algorithm and its variations will be DNA sequence and Amino Acid sequences. Input processing unit will process sequences three at a time as three neighborhood cellular automata is considered for processing DNA sequences. The rule generator will transform the complemented and non complemented rules in the form of matrix, so that we can apply the rules to the corresponding sequence positions very easily. AIS-MACA-Z basins are calculated as per the instructions of proposed algorithm and an inverter tree named as AIS multiple attractor cellular automata is formed which can predict the class of the input after all iterations.

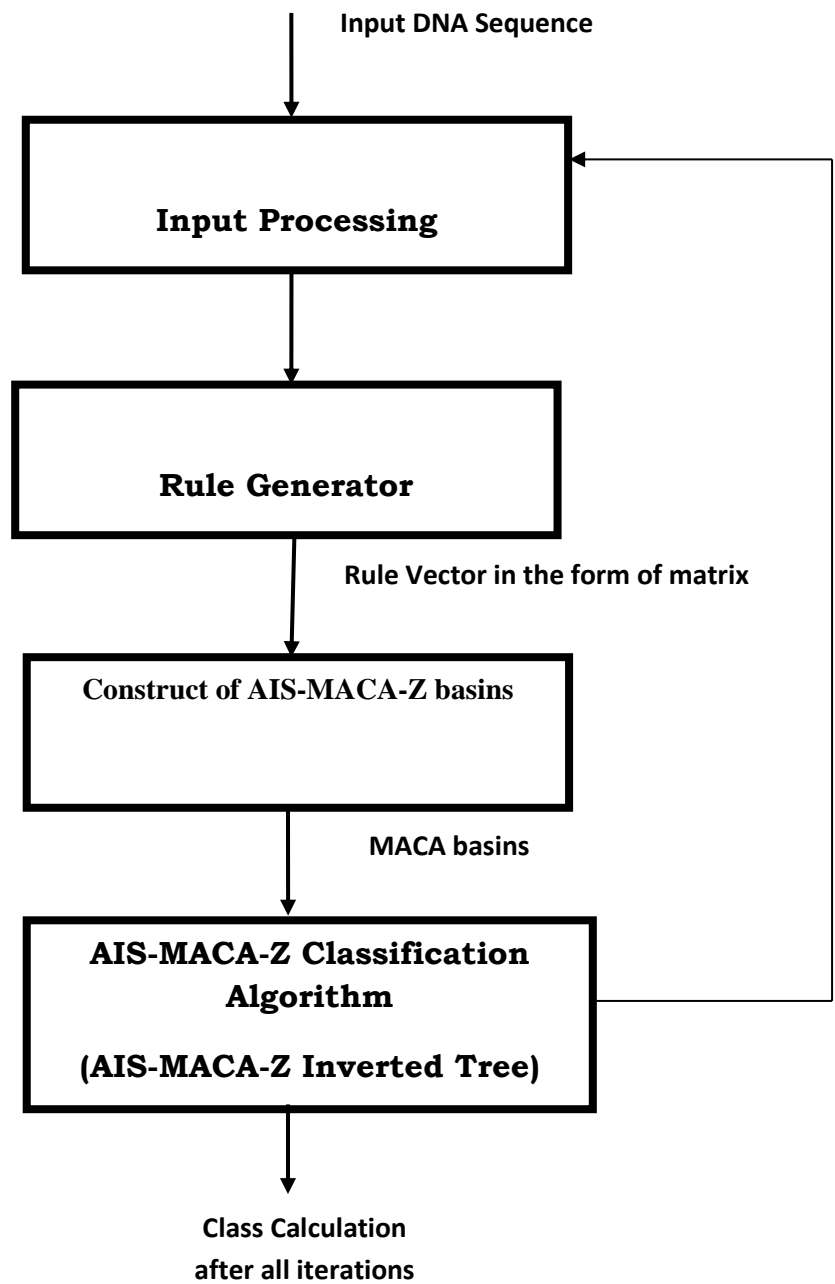

*Fig. 1:* *General Architecture of AIS-MACA- Z.*

For a sample DNA sequence and fuzzy real values, the data structures AIS-MACA-Z [7,8] is shown in the figure 2.

The decimal equivalent of the next state function, as defined as the rule number of the CA cell. In a 2-state 3-neighborhood CA, there are 256 distinct next state functions, among 256 rules, rule 51is represented in the following Eqn. (1).

Rule 51 : $q_i(t + 1) = q_i(t)$ (1)

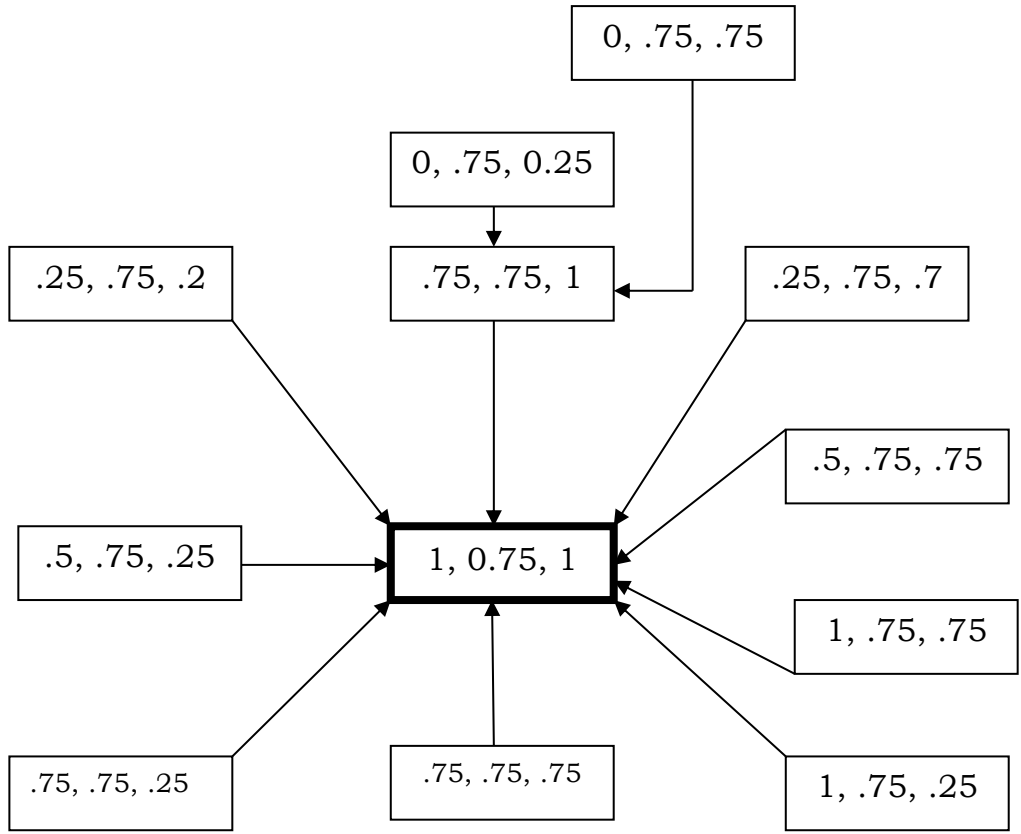

*Fig. 2:* AIS-MACA- Z Data Structure.

**Experimental Results**

Experiments were conducted by using Fickett and Toung data [9] for predicting the protein coding regions and splicing cites. All the 21 measures reported in [9] were considered for developing the classifier. For promoter region identification human promoters from EPDnew[10]. Table 1 represents the splicing cite output. Figure 3,4,5,6 shows the prediction of promoter and protein coding regions.

*Table 1:* Splicing Cite Output.

| G Str | Feature | Start | End | Score |
|---|---|---|---|---|
| 1 + | 1 CDSf | 417 - | 491 | 8.98 |
| 1 + | 2 CDSi | 942 - | 1000 | 0.61 |
| 1 + | 3 CDSi | 2072 - | 2223 | 24.72 |
| 1 + | 4 CDSi | 2328 - | 2453 | 14.03 |
| 1 + | 5 CDSi | 2543 - | 2640 | 3.22 |
| 1 + | 6 CDSi | 3857 - | 3983 | 20.83 |
| 1 + | 7 CDSi | 4079 - | 4173 | 12.55 |

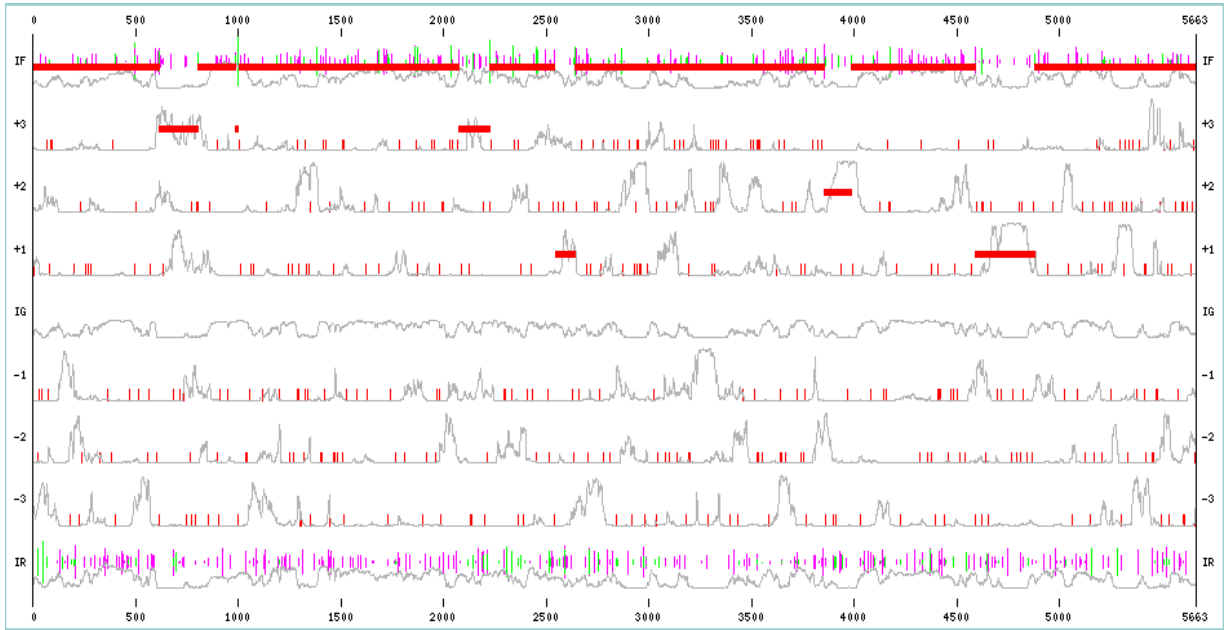

*Fig. 3:* *AIS-MACA- Z Interface Identifying Protein Coding Regions.*

| Gene number | Element number | Exons/UTR | Strand | Left end | Right end | Length | Phase | Frame |
|---|---|---|---|---|---|---|---|---|
| 1 | 1 | Internal | + | 615 | 801 | 187 | +1 | +3 |
| 1 | 2 | Internal | + | 985 | 1000 | 16 | +2 | +3 |
| 1 | 3 | Internal | + | 2072 | 2223 | 152 | +3 | +3 |
| 1 | 4 | Internal | + | 2543 | 2640 | 98 | +2 | +1 |
| 1 | 5 | Internal | + | 3857 | 3983 | 127 | +1 | +2 |
| 1 | 6 | Internal | + | 4589 | 4879 | 291 | +2 | +1 |

*Fig. 4:* *Exons Boundary Reporting.*

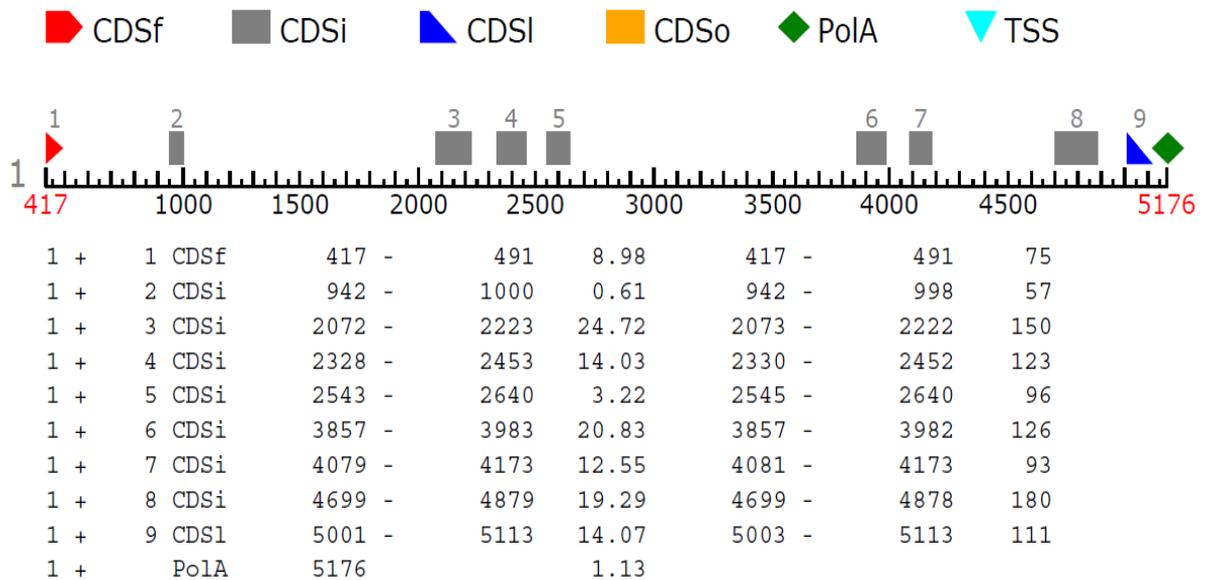

*Fig. 5:* *Coding Sequence Reporting.*

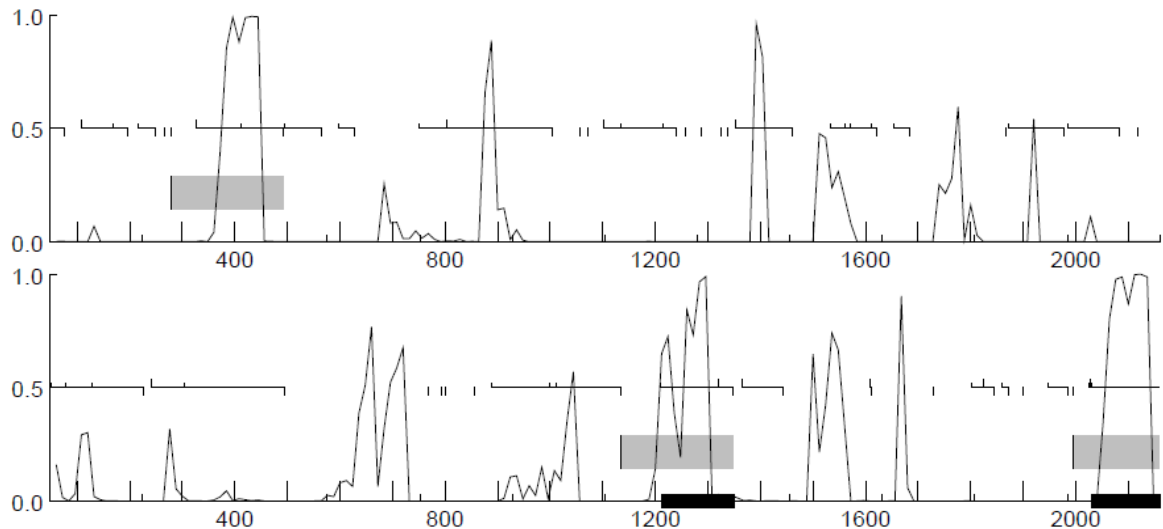
*Fig. 6: Coding Sequence Probability Levels.*

## CONCLUSION

We have developed a logical classifier designed with MACA and strengthened with AIS technique that uses a fuzzy logic for predicting the slicing sites, protein and promoter regions. The accuracy of the AIS-MACA-Z classifier is considerably more when compared with the existing algorithm which is 90.6% in average. The proposed classifier can handle large data sets and sequences of various lengths. This classifier certainly provides intuition towards application of MACA to several problems in bioinformatics.

## REFERENCES

1. Bevilacqua, Vitoantonio, Maurizio Triggiani et al. An expert system for an innovative discrimination tool of commercial table grapes. In *Intelligent Computing Theories and Applications*, Springer Berlin Heidelberg, 2012; 7390: 95–102p.
2. Ganguly, Niloy, Biplab K. Sikdar, et al. A survey on cellular automata. (2003).
3. Sarkar, Palash, and Subhamoy Maitra, Nonlinearity bounds and constructions of resilient Boolean functions. In *Advances in Cryptology-CRYPTO 2000,* Springer Berlin Heidelberg, 2000; 1880: 515–532p.
4. Maji, Pradipta, Chandrama Shaw et al. Theory and application of cellular automata for pattern classification. In *Fundamenta Informaticae*, 2003; 58(3): 321–354p.
5. Yin, Changchuan, and Stephen S-T. Yau, Prediction of protein coding regions by the 3-base periodicity analysis of a DNA sequence. In *Journal of theoretical biology*, 2007;247(4): 687–694p.
6. Mena-Chalco, Jesús P., Helaine Carrer et al. Identification of protein coding regions using the modified Gabor-wavelet transform. In *Computational Biology and Bioinformatics,* IEEE/ACM Transactions on 2008; 5 (2): 198–207p.
7. Sree Pokkuluri Kiran, AIS-INMACA: A Novel Integrated MACA Based Clonal Classifier for Protein Coding and Promoter Region Prediction. In *J Bioinfo Comp Genom*, 2014; 1: 1–7p.
8. Nedunuri, SSSN Usha Devi, Inampudi Ramesh Babu, and Pokkuluri Kiran Sree, An Extensive Repot on Cellular Automata Based Artificial Immune System for Strengthening Automated Protein Prediction. In *Advances in Biomedical Engineering Research*, 2013;1(3): 1310–4342p.
9. Fickett, James W., and Chang-Shung Tung, Assessment of protein coding measures. In *Nucleic acids research,*1992; 20 (24): 6441–6450p.
10. Dreos, René, Giovanna Ambrosini et al. EPD and EPDnew, high-quality promoter resources in the next-generation sequencing era. In *Nucleic acids research*, 2013; 41(D1): D157–D164p.